\documentclass[a4paper]{jpconf}
\usepackage{graphicx}
\usepackage{mathtools}
\usepackage{amssymb}
\usepackage{bbold}
\usepackage{hyperref}
\DeclareMathAlphabet{\mathpzc}{OT1}{pzc}{m}{it}

\DeclarePairedDelimiter\bra{\langle}{\rvert}
\DeclarePairedDelimiter\ket{\lvert}{\rangle}

\DeclarePairedDelimiterX\braket[2]{\langle}{\rangle}{#1 \delimsize\vert #2}

\begin{document}

\title[Spatial Phase Separation of a Binary Mixture in a Ring Trimer]{Spatial Phase Separation of a Binary Mixture in a Ring Trimer}

\author{Vittorio Penna and Andrea Richaud}
\address{ Dipartimento di Scienza Applicata e Tecnologia and u.d.r. CNISM, Politecnico di Torino, 
Corso Duca degli Abruzzi 24, I-10129 Torino, Italy}
\ead{andrea.richaud@polito.it}

\begin{abstract}
We investigate the phase separation mechanism of bosonic binary mixtures in spatially-fragmented traps, evidencing the emergence of phases featuring a different degree of mixing. The analysis is initially carried out by means of a semiclassical approach which transparently shows the occurrence of critical phenomena. These predictions are actually corroborated by the study of genuinely quantum indicators, including, but not limited to, the energy levels' structure and the entanglement between the species. The scope of our work goes also beyond the ground state's properties, as it comprises excited states and the dynamical evolution thereof. In particular, after introducing an indicator to monitor the degree of mixing, we show that several dynamical regimes feature persistent demixing in spite of their remarkably chaotic character.

\end{abstract}

\section{Introduction}
The localization of two interacting condensed species in different spatial regions is a phenomenon that occurs in presence of strong interspecies repulsive couplings. The mechanism underpinning this spatial phase separation has been deeply investigated in systems of ultracold bosons within the mean-field approach to condensate's dynamics \cite{cmixt3,cmixt4,cmixt5}. It has been evidenced that a number of elements, including the depth of the trapping potential, the number of particles and the intraspecies interactions, influence the emergence of the mixing-demixing transition and the way it develops \cite{Viktor}. From the experimental side, the unprecedented control of quantum matter has allowed to realize multicomponent \cite{Zenesini, Inguscio,Gadway,Soltan} and spatially-fragmented \cite{jz,bdz,yuk} condensed systems whose rich phenomenology, in turn, rekindled the interest of the theoretical community for a deep comprehension of the demixing transition in the presence of optical lattices \cite{Belemuk, Bruno}. 

Systems made up of two condensed species (be they heteronuclear \cite{Catani_deg} or homonuclear \cite{Soltan_2}) indeed offer a plethora of observable physical phenomena ranging from the dynamics of solitons \cite{Makarov,Ivan,Gallemi_correnti} to the formation of entanglement between the species \cite{ent,NoiEntropy}, not to mention the mechanisms that directly account for phase separation \cite{sep2,sep3, Angom}. 

In this context, the present manuscript aims to discuss the phenomenon of spatial phase separation in the two simplest possible geometries, the dimer and the trimer, whose fragmented character strongly affects the mixing properties of the bosonic binary mixture. Such systems are thoroughly analyzed in section \ref{sec:Dimero} and \ref{sec:trimero} respectively. In both sections we start presenting the Bose-Hubbard (BH) Hamiltonian, the second-quantized model which well captures the essence of the physical problem. We then switch to a semiclassical perspective, as we think that it allows for the emergence of the mixing-demixing transition at its most transparent. We support the predictions provided by the aforementioned approximation scheme by introducing genuinely quantum indicators. The latter, which include the energy spectrum and the entanglement entropy (EE), prove to be singular where anticipated by the semiclassical approach. Eventually, in section \ref{sec:chaos}, we extend our analysis beyond the ground-state physics by providing an overview on the interplay between chaotic dynamics and persistent demixing.


\section{The binary mixture in a double well}
\label{sec:Dimero}
In order to effectively model a system of two condensed species in a two-well potential, we introduce operators $a_L$, $a_R$, $b_L$ and $b_R$.  Operator $a_L$ $(a_R)$ destroys a species-a boson in the left (right) well. Similarly, operator $b_L$ $(b_R)$ destroys a species-b boson in the left (right) well. Standard bosonic commutators hold, namely $[a_X,a_Y^\dagger]=\delta_{X,Y}=[b_X,b_Y^\dagger]$ and $[a_X,b_Y]= $ $[a_X,b_Y^\dagger]=0$, with $X,Y \, \in \{R,L\}$. In this framework, the BH Hamiltonian associated to the system reads
$$
  \hat{H}= -T_a \left(a_L a_R^\dagger +a_R a_L^\dagger  \right) + \frac{U_a}{2} \left[n_L(n_L-1)+ n_R(n_R-1)  \right]
$$
\begin{equation}
    \label{eq:Hami_dimero}
  -T_b \left(b_L b_R^\dagger + b_R b_L^\dagger  \right) + \frac{U_b}{2} \left[m_L(m_L-1)+ m_R(m_R-1)  \right]
+ W\left(n_L m_L+ n_R m_R \right),
\end{equation}
where $n_X=a_X^\dagger a_X$ ($m_X=b_X^\dagger b_X$) is the numbers of species-a (species-b) bosons in well $X\in\{R,L\}$.  This model includes tunnelling terms $T_a$ and $T_b$, intraspecies interactions $U_a$ and $U_b$, and a density-density interspecies coupling $W$. The boson number of each atomic species, namely $N:=n_L+n_R$ and $M=m_L+m_R$ represent two independent conserved quantities, being $[N,\hat{H}]=[M,\hat{H}]=0$. 

\subsection{CVP approach to the mixing-demixing transition}
\label{sec:CVP_Dimero}
\begin{figure}
    \centering
    \includegraphics[width=1\columnwidth]{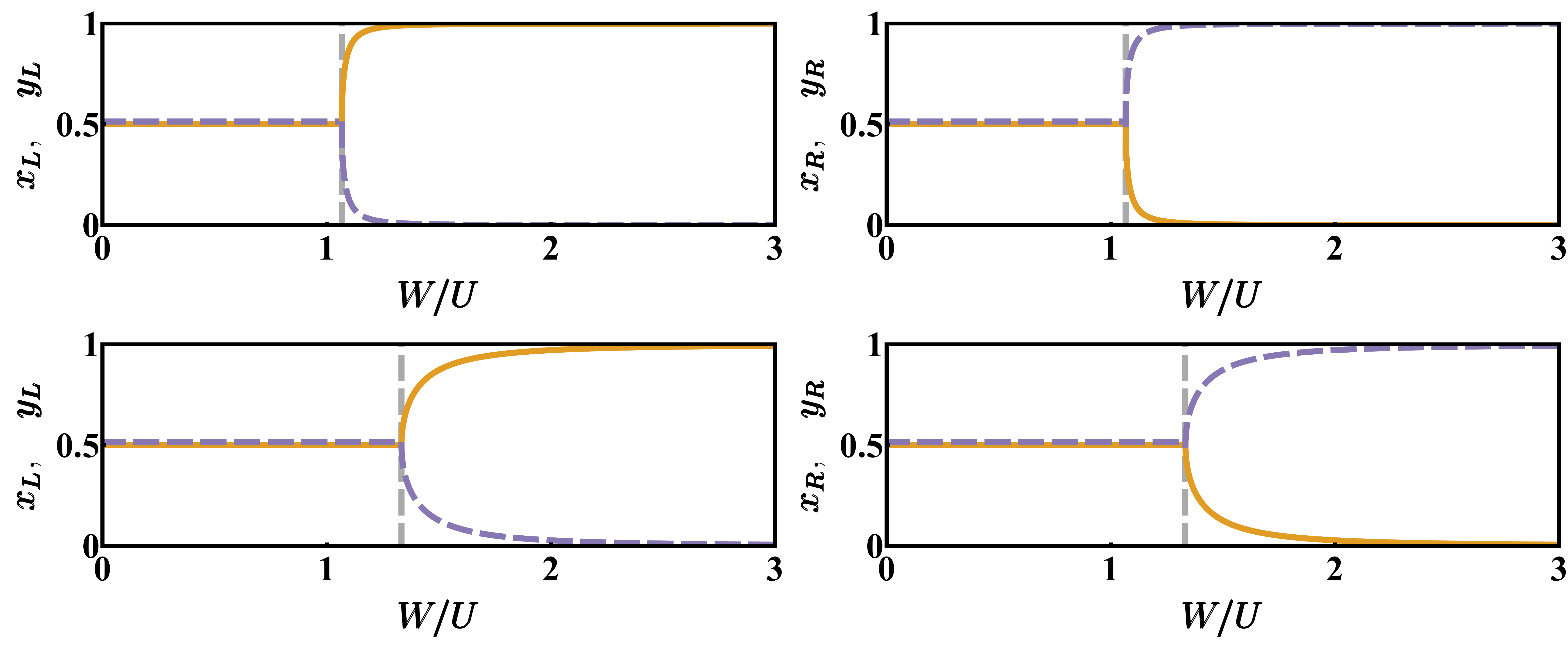}
    \caption{Boson populations in the two wells as functions of $W/U$. Solid orange (purple dashed) line corresponds to species-a (species-b) bosons. Upper row refers to $T=1$, while the second row has been drawn for $T=5$. In all panels, parameters $N=30$ and $U=1$ have been chosen. Vertical gray lines correspond to the limiting condition of inequality (\ref{eq:Critical_Dimer}).}
    \label{fig:Dimero_CVP}
\end{figure}

In order to investigate the miscibility properties of the ground state of Hamiltonian (\ref{eq:Hami_dimero}), we switch to a description of boson populations in terms of continuous variables $x_X:=n_X/N$ and $y_X:=m_X/M$, with $X\in\{L,R\}$. This technique, successfully used in the study of spatially-fragmented BECs \cite{Spekkens, PennaLinguaPRE, Ciobanu, NoiSREP}, allows one to recast the computation of the ground state of quantum Hamiltonian (\ref{eq:Hami_dimero}) into the search for the global minimum of the classical effective potential 
$$
  V_{dimer} = -2NT_a\sqrt{x_L x_R} - 2MT_b \sqrt{y_L y_R} + \frac{U_aN^2}{2} \left(x_L^2+x_R^2\right) +   \frac{U_bM^2}{2} \left(y_L^2+y_R^2\right) 
$$
\begin{equation}
\label{eq:V_dimer}
     + WNM\left(x_L y_L + x_R y_R\right)
\end{equation}
In the following, we will assume, for simplicity, $T_a=T_b=:T$, $U_a=U_b=:U$ and $N=M$ but we note in advance that small deviations from this ideal scenario do not substantially affect our analysis. As the intuition suggests, if the interspecies repulsion $W$ is small compared to the intraspecies repulsion $U$, than the two species are mixed and uniformly distributed in the two wells. More specifically, one can verify that the uniform configuration $x_X=y_X=1/2$ (with $X\in\{L,R\}$) is indeed a global minimum of effective potential (\ref{eq:V_dimer}) provided that \begin{equation}
\label{eq:Critical_Dimer}
   \frac{W}{U}< 1 + \frac{2T}{UN}.
\end{equation}
Above this critical value, the two species start to separate in different spatial regions, i.e. to localize in different wells. The analytic expressions of boson populations in the two wells,
\begin{equation}
\label{eq:x_L_dimer}
    x_L=\frac{1}{2}+\frac{1}{2}\sqrt{1-\left[\frac{2T}{N(W-U)}\right]^2}, \qquad x_R=1-x_L
\end{equation}
\begin{equation}
\label{eq:y_L_dimer}
    y_L=\frac{1}{2}-\frac{1}{2}\sqrt{1-\left[\frac{2T}{N(W-U)}\right]^2}, \qquad y_R=1-y_L,
\end{equation}
show that, for $W\to+\infty$, the two species completely demix, meaning that species-a bosons occupy just the left well while species-b bosons localize just in the right well (namely, $x_L=1$, $x_R=0$, $y_L=0$, $y_R=1$). 
We remark that, due to the system symmetry, formulas (\ref{eq:x_L_dimer}) and (\ref{eq:y_L_dimer}) represent just one of the two minimum-energy configurations, the other one being obtained swapping labels ``$L$" and ``$R$". It is interesting to notice (see Fig. \ref{fig:Dimero_CVP}) that, all other things being equal, the bigger the tunnelling amplitude $T$, the bigger the interspecies repulsion $W$ needed to trigger species demixing, the smoother the transition from the fully mixed to the completely demixed configuration. 

\subsection{Spectrum collapse at the critical point} 
\begin{figure}
    \centering
    \includegraphics[width=0.5\columnwidth]{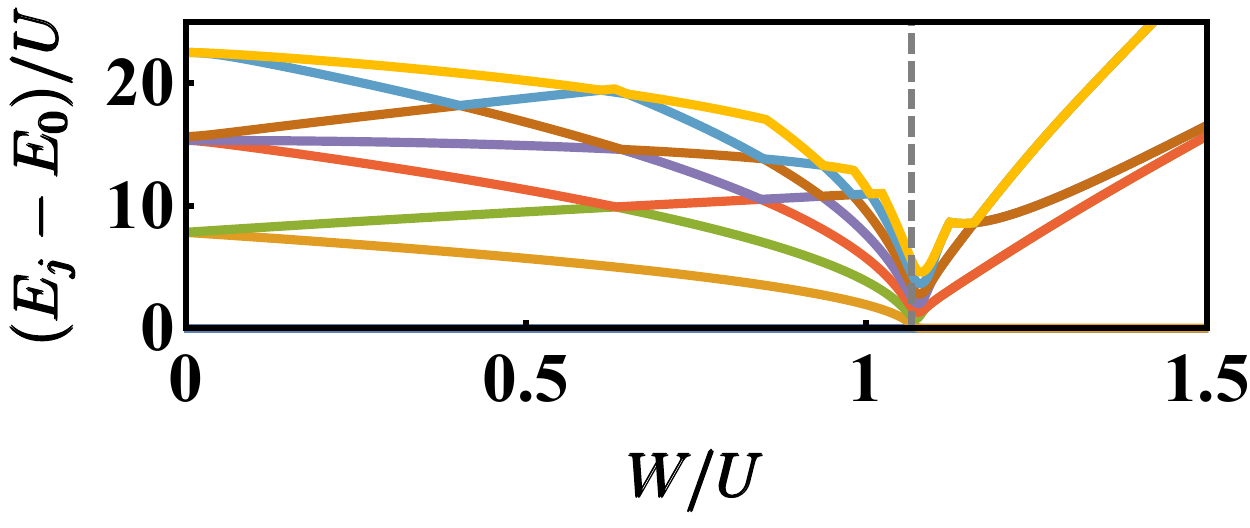}
    \caption{Energy spectrum (first $9$ levels) as a function of $W/U$. Model parameters $N=30$, $U=1$, $T=1$ have been chosen. Vertical gray line correspond to the limiting condition of inequality (\ref{eq:Critical_Dimer}).}
    \label{fig:Dimer_energy_levels}
\end{figure}

The energy level distribution considerably varies according to the ratio $W/U$. As depicted in Fig. \ref{fig:Dimer_energy_levels}, in fact, the energy levels' structure is very different at the two sides of the critical point and exhibits a collapse where the mixing-demixing transition occurs.
This circumstance, evidenced by means of an exact numerical diagonalization of Hamiltonian (\ref{eq:Hami_dimero}), can be better highlighted if one resorts to the Bogoliubov technique for the computation of quasi-particles frequencies. Within this approximation scheme \cite{PennaLinguaPRE}, and in agreement with the results of the CVP approach, characteristic frequencies
$$
     \omega_{1,2} = \sqrt{T\left(T+N\frac{U\pm W}{2}\right)}   
$$
are well defined, in fact, only in that region of parameters' space identified by inequality (\ref{eq:Critical_Dimer}).

\subsection{Entanglement entropy as a critical indicator} 
\begin{figure}
    \centering
    \includegraphics[width=1\columnwidth]{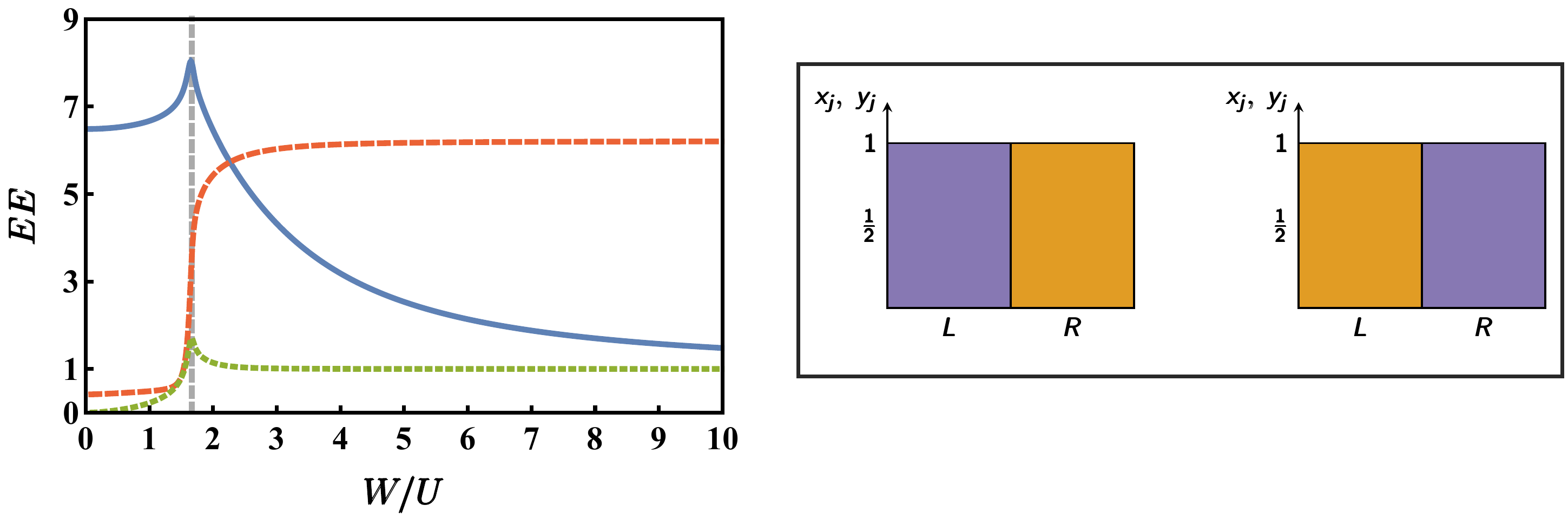}
    \caption{\textbf{Left panel:} Entanglement Entropy as a function of $W/U$ for three different partition schemes: blue solid line corresponds to site-mode partition (L-R); red dashed line to momentum-mode partition (S-D); green dotted line to species-mode partition (a-b). Model parameters $T=1$, $U=0.1$, $N=M=30$ have been chosen. Vertical gray line corresponds to critical condition of inequality (\ref{eq:Critical_Dimer}) which, in the present case, reads $(W/U)_c \approx 1.67$. \textbf{Right panel:} pictorial representation of the system's ground state for $W/U\gg 1$. In this circumstance, it corresponds to the quantum superposition of two semiclassical configurations and exhibits the Schr\"{o}dinger-cat like structure defined by equation (\ref{eq:Dimero_CAT}).}
    \label{fig:Dimero_EEs}
\end{figure}

The study of the correlation properties between two partitions of the system's Hilbert space has proved to be an effective indicator for detecting the occurrence of mixing-demixing transitions \cite{NoiEntropy, NoiSREP}. The Entanglement Entropy (EE) is the indicator commonly used to quantify these quantum correlations and the standard way to compute it involves the subdivision of the space of states into two partitions, $V_1$ and $V_2$. More specifically, the entanglement between $V_1$ and $V_2$ is given by 
\begin{equation}
        EE_{V_1-V_2} = - \mathrm{Tr}_{V_1} (\hat{\rho}_{V_1}\, \log_2 \hat{\rho}_{V_1} )
\label{eq:EE_generale}
\end{equation}
and corresponds to the Von Neumann entropy of the reduced density matrix 
\begin{equation}
     \hat{\rho}_{V_1}=\mathrm{Tr}_{V_2} \left(\hat{\rho}_0\right)
\label{eq:Reduced_density_matrix}
\end{equation}
which is obtained, in turn, tracing out the degrees of freedom of $V_2$ from the ground state's density matrix $\hat{\rho}_0=\ket{\psi_0}\bra{\psi_0}$. We remark that the partitioning-step is a crucial one and that a strong entanglement in a certain partition may correspond to a weak (or null) entanglement in another one. In \cite{NoiEntropy}, we have investigated three different kinds of EE according to three different ways of partitioning the two-species-dimer system: \textit{i)} $EE_{L-R}$, between the site modes; \textit{ii)} $EE_{s-d}$ between momentum modes and \textit{iii)}  $EE_{a-b}$ between species modes. Interestingly, we have evidenced that they build up, respectively, due to the presence of a hopping amplitude $T$ between the sites, because of an on-site repulsion $U$ and owing to the interspecies coupling $W$. On top of that, as clearly illustrated in Fig. \ref{fig:Dimero_EEs}, all three indicators feature singularities (either peaks or jumps) where the mxing-demixing transition takes place.  
Moreover, one can notice that, for large interspecies interactions, the entanglement between the species $(EE_{a-b})$ tends to the limiting value $1=\log_2 2$, a number which is reminiscent of the \textit{two}-sided structure of the system's ground-state. In that circumstance, in fact, the latter, can be well approximated by the Schr\"{o}dinger-cat-like state
\begin{equation}
\label{eq:Dimero_CAT}
    \ket{\psi_0}= \frac{1}{\sqrt{2}} \left( \ket{N,0}_L \ket{0,N}_R + \ket{0,N}_L \ket{N,0}_R \right)
\end{equation}
where state $\ket{n,m}_L\ket{N-n,N-m}_R$ is such that there are $n$ species-a and $m$ species-b bosons in the left well.

The concept of EE, which is a quantity strictly defined for pure states, can be readily generalized to non-zero temperatures by substituting $\hat{\rho}_0$ with the thermal density matrix
$$
   \hat{\rho}_T = \frac{1}{\mathcal{Z}}\sum_{n}e^{-\beta E_n} \ket{\psi_n}\bra{\psi_n}, \qquad \text{where}\quad \mathcal{Z}=\sum_n e^{-\beta E_n}     \quad \text{and} \quad \hat{H}\ket{\psi_n}=E_n\ket{\psi_n} 
$$
in formulas (\ref{eq:EE_generale}) and (\ref{eq:Reduced_density_matrix}). This indicator, which goes under the name of bipartite residual entropy, despite being affected by classical correlations between the subsystems and by the classical entropy, is still able to detect the occurrence of mixing-demixing transitions when these occur at finite temperature \cite{NoiEntropy}. 

\section{The binary mixture in a ring trimer}
\label{sec:trimero}
We turn to the analysis of the mixing properties of a bosonic binary mixture in a ring trimer, a system where the presence of an additional well discloses a phenomenology much richer than the one discussed in Section \ref{sec:Dimero}. Similarly to what discussed in relation to the two-species dimer, we start with introducing the second-quantized operator $a_i$ ($b_i$) which annihilates a species-a (species-b) particle in the $i-$th well (where $i=1,2,3$) and whose bosonic character is expressed by the following commutation relations: $[a_i,a_j^\dagger]=\delta_{i,j}$, $[b_i,b_j^\dagger]=\delta_{i,j}$. Within this formalism, the model of a binary mixture in a three-well potential with periodic boundary conditions is effectively portrayed by two BH Hamiltonians (each one being associated to a bosonic species and involving three spatial modes) coupled by an interspecies repulsion term. This reads:
$$
   \hat{H}=\frac{U_a}{2}\sum_{j=1}^3 n_j (n_j-1) -T_a\sum_{j=1}^3\left(a_{j+1}^\dagger a_j +a_j^\dagger a_{j+1}\right ) + \frac{U_b}{2}\sum_{j=1}^3 m_j (m_j-1)  -T_b\sum_{j=1}^3\left(b_{j+1}^\dagger b_j +b_j^\dagger b_{j+1}\right ) 
$$
\begin{equation}
\label{eq:Trimer_Hami}
   + W\sum_{j=1}^3 n_j m_j 
\end{equation}
where the periodic boundary conditions implied by the ring geometry are such that $j=4 \equiv 1$. After recalling that $n_j=a_j^\dagger a_j$ and $m_j=b_j^\dagger b_j$, we remark that the total number of bosons in the two species, $N=\sum_j n_j$ and $M=\sum_j m_j$, constitute, as in the dimer case, two independent conserved quantities. Terms proportional to the hopping amplitudes $T_a$ and $T_b$ favour particles' delocalization in the ring geometry; conversely, terms proportional to intraspecies repulsive interactions $U_a$ and $U_b$ favor particles' localization. Eventually, the interspecies-repulsion term ($\propto W$) is responsible for the spatial phase separation of the two bosonic species.

\subsection{CVP approach to the two-step demixing transition}
\label{sec:CVP_Trimero}
As explained in Section \ref{sec:CVP_Dimero}, an effective way to investigate the phase separation mechanism is to switch to the CVP, an approximation scheme where the inherently discrete eigenvalues of number operators $n_j$ and $m_j$ are replaced by continuous variables $x_j=n_j/N$ and $y_j=m_j/M$. In this framework, the ground state of quantum Hamiltonian (\ref{eq:Trimer_Hami}) corresponds to the global minimum of classical potential 
$$
  V_{trimer} =-2NT_a\left(\sqrt{x_1x_2}+\sqrt{x_2x_3}+\sqrt{x_3x_1}\right) + \frac{U_aN^2}{2}\left(x_1^2+x_2^2+x_3^2\right) 
$$
$$
 -2MT_b\left(\sqrt{y_1y_2}+\sqrt{y_2y_3}+\sqrt{y_3y_1}\right) + \frac{U_bM^2}{2}\left(y_1^2+y_2^2+y_3^2\right) 
$$
\begin{equation}
\label{eq:V_trimer}
     + W(x_1y_1+x_2y_2+x_3y_3)
\end{equation}
subject to the constraints $0<x_j<1$, $0<y_j<1$ $\forall j$, $\sum_{j}x_j=\sum_{j}y_j=1$. For the sake of simplicity, in the following we will assume that $U_a=U_b=:U$, $T_a=T_b=:T$ and $N=M$, although small offsets from this ideal condition do not substantially affect the results. 

The basic difference with respect to the two-species dimer is that, in this more complex geometry, tuning the ratio $W/U$ one triggers \textit{two} mixing-demixing transitions, rather than only one. In other words, one can recognize three different phases, which differ in the degree of mixing: 
\begin{enumerate}
    \item Fully mixed phase. If $W/U \in \left(0,\, 1+\frac{9T}{2UN}\right)$ then the two species are uniformly distributed in the ring trimer, meaning that $x_j=y_j=1/3\, \forall j$. 
    \item Partially (de)mixed phase. If $W/U \in \left(1+\frac{9T}{2UN}, \, R_c \right)$, where $R_c=R_c(T,U,N)>2$ then one can recognize a rather low degree of mixing in two out of three wells and an almost perfect  mixing in the remaining one. It is worth noticing that, in analogy with the two-species dimer (see section \ref{sec:CVP_Dimero}), increasing the ratio $T/(UN)$ has the twofold effect of shifting critical points $1+\frac{2T}{UN}$ and $R_c$ and of smoothing the transitions.  
    \item Fully demixed phase. If $W/U>R_c$ then the two species completely separate, meaning that one species occupies two wells while the second species conglomerates in the remaining site.
\end{enumerate}
\begin{figure}
    \centering
    \includegraphics[width=1\columnwidth]{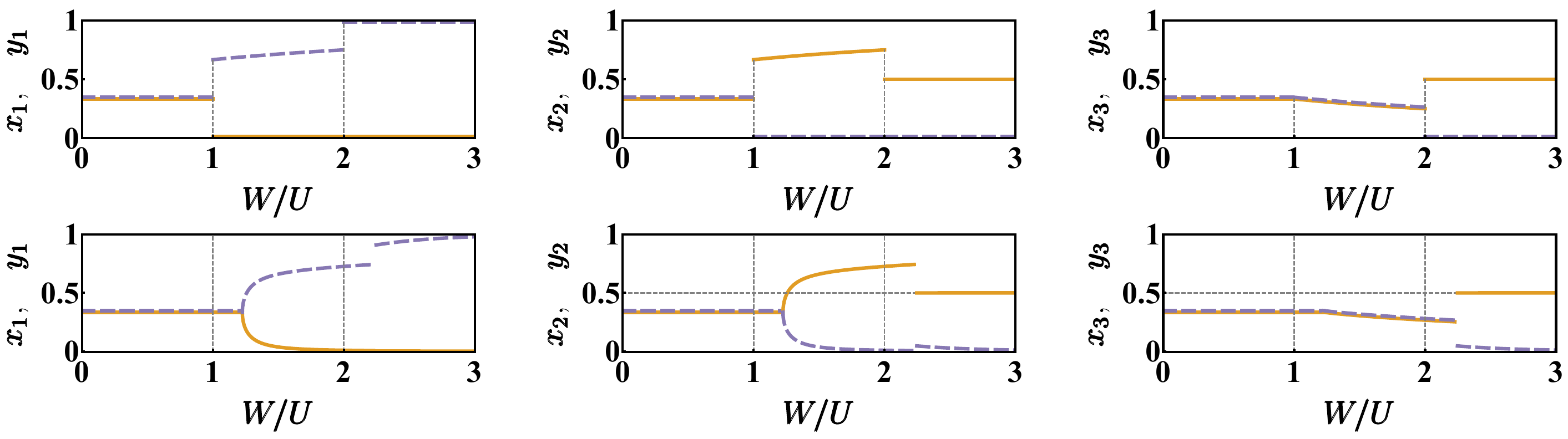}
    \caption{Boson populations in the three wells as functions of $W/U$. Solid orange (purple dashed) line corresponds to species-a (species-b) bosons. Upper row refers to $T=0$, while the second row has been drawn for $T=1.25$. In all panels, parameters $N=50$ and $U=1$ have been chosen.  }
    \label{fig:Trimero_CVP}
\end{figure}

This scenario, illustrated in the second row of Fig. \ref{fig:Trimero_CVP}, turns even clearer and sharper if one considers the semiclassical limit $N \to +\infty$ (or, equivalently, the no-tunnelling limit $T\to 0$). In this circumstance, in fact, a fully-analytic investigation of the two-step mixing-demixing transition can be carried out \cite{NoiSREP}. Such analysis is rather complex because the non-convex character of effective potential (\ref{eq:V_trimer}) makes it necessary an exhaustive exploration of its domain boundaries. The result is that the partially (de)mixed phase occurs for $W/U\in(1,2)$, a range where boson populations behave as shown table \ref{tab:Popolazioni_fase_2}.

\begin{table}[ht]
\label{tab:Popolazioni_fase_2}
    \centering
    \begin{tabular}{ |c c c| c c c| c c c|}
        \hline
        $x_1=0$ & & & & $x_2=\frac{1+\frac{W}{U}}{2+\frac{W}{U}}$ & & & & $x_3=\frac{1}{2+\frac{W}{U}}$ \\
        \hline
        $y_1=\frac{1+\frac{W}{U}}{2+\frac{W}{U}}$ & & & & $y_2=0$ & & & & $y_3=\frac{1}{2+\frac{W}{U}}$ \\
        \hline
\end{tabular}
\caption{Boson populations in the partially (de)mixed phase, in the limit $N\to+\infty$.}
\end{table}
We observe that the aforementioned configuration, depicted in the first row of Fig. \ref{fig:Trimero_CVP}, is not the only one which minimizes effective potential (\ref{eq:V_trimer}). The latter, in fact, owing to the system's symmetry, features three isoenergetic points of minimum which are equal up to cyclic permutations of the site's indexes. The link between symmetry and number of minimum-energy configurations will be made clearer in section \ref{sec:Trimero_EE}.

\subsection{Spectrum collapse at the mixing-demixing transitions}
\begin{figure}
    \centering
    \includegraphics[width=0.5\columnwidth]{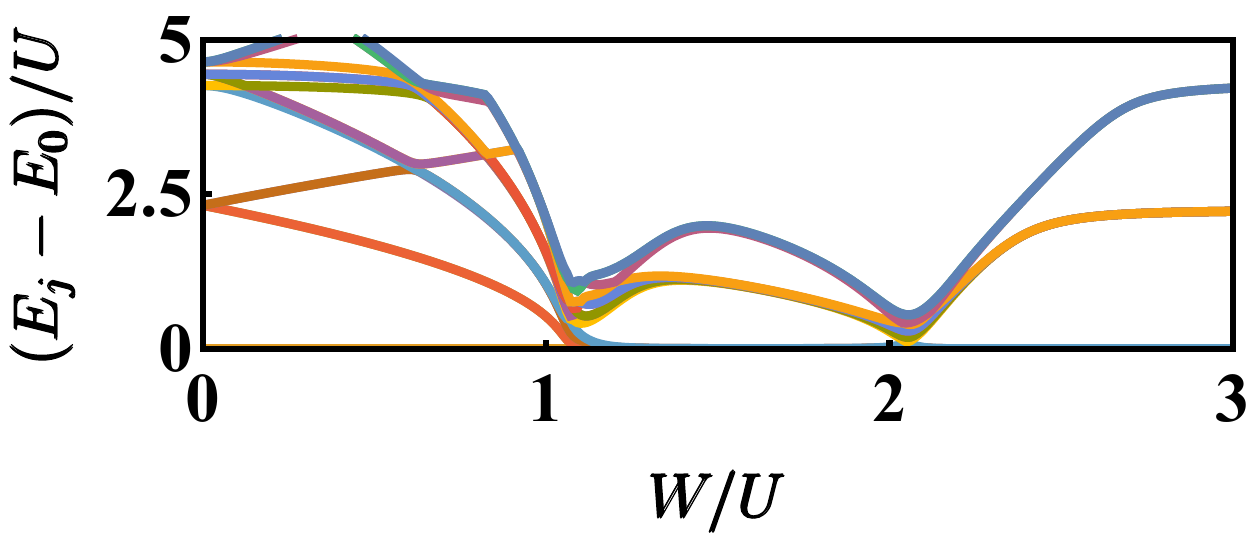}
    \caption{Energy spectrum (first $15$ levels) as a function of $W/U$. Model parameters $N=15$, $U=1$, $T=0.2$ have been chosen.}
    \label{fig:Trimer_energy_levels}
\end{figure}

We have performed an exact numerical calculation of the energy levels of Hamiltonian (\ref{eq:Trimer_Hami}) and the result, shown in Fig. \ref{fig:Trimer_energy_levels}, clearly illustrates the spectrum collapse and reorganization across the two critical points.
In this regard, it is worth noticing that, the smaller the ratio $T/(UN)$, the sharper the collapse, the more abrupt the transition.

A deeper physical insight can be gained from the study of quasi-particle frequencies which feature the fully mixed phase. Within the Bogoliubov approximation scheme and making use of the dynamical algebra method, we were able to determine such frequencies for a ring made up of a generic number of sites \cite{NoiPRA2}. In the specific case of a ring trimer, their explicit expression
$$
   \omega_{1,2}= \sqrt{3T\left(3T+\frac{2UN}{3} \pm\frac{2NW}{3} \right)}
$$
shows that they are well defined only if 
$$
   \frac{W}{U}<1+\frac{9T}{2UN}
$$
an equality constituting the same condition already emerged in the context of the CVP approach (see section \ref{sec:CVP_Trimero}).

\subsection{Entanglement between the species}
\label{sec:Trimero_EE}
Among the different partition schemes which can be applied to the Hilbert space of states associated to Hamiltonian (\ref{eq:Trimer_Hami}), we focus on the one involving species-a and species-b modes. To be more specific, after numerically computing the ground state $\ket{\psi_0}$ of (\ref{eq:Trimer_Hami}) and the relevant density matrix $\hat{\rho}_0=\ket{\psi_0}\bra{\psi_0}$, one obtains the reduced density matrix  $\hat{\rho}_a=\mathrm{Tr}_b\left(\hat{\rho}_0\right)$ associated to species-a bosons by tracing away the degrees of freedom relevant to species-b bosons. The entanglement between the species can be thus computed as
\begin{equation}
\label{eq:Trimero_EE_gatto}
       EE_{a-b} = -\mathrm{Tr}_a \left(\hat{\rho}_a \log_2  \hat{\rho}_a \right).
\end{equation}
\begin{figure}
    \centering
    \includegraphics[width=1\columnwidth]{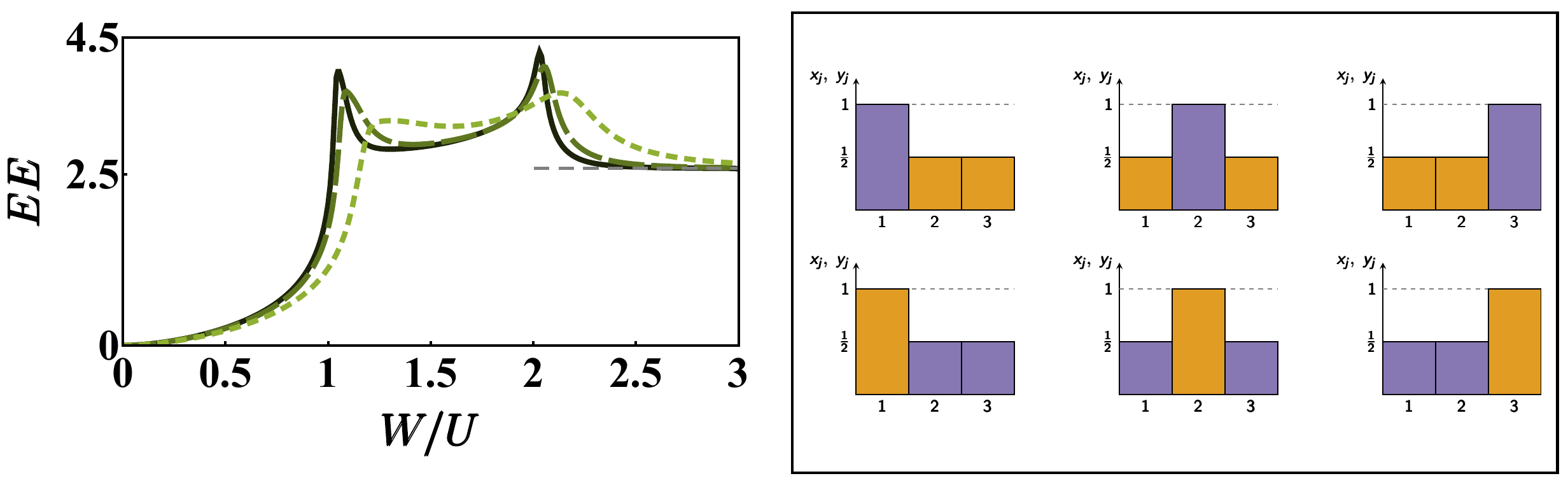}
    \caption{\textbf{Left panel:} Entanglement Entropy as a function of $W/U$ for three different hopping amplitudes: solid line corresponds to $T=0.12$, dashed line to $T=0.20$, dotted line to $T=0.50$. Model parameters $U=1$, $N=M=15$ have been chosen. Horizontal gray line corresponds to the limiting value $\log_2(6)$. \textbf{Right panel:} pictorial representation of the system's ground state for $W/U\gg 2$. In this circumstance, it corresponds to the quantum superposition of six semiclassical configurations and exhibits a Schr\"{o}dinger-cat like structure.}
    \label{fig:Trimero_EE}
\end{figure}
With reference to the left panel of Fig. \ref{fig:Trimero_EE}, where we illustrate how indicator (\ref{eq:Trimero_EE_gatto}) changes as a function of the ratio $W/U$, we can notice that the entanglement between the species: \textit{i)} is correctly null when they do not interact, i.e. when $W=0$; \textit{ii)} features peaks where the mixing-demixing transitions occur; \textit{iii)} tends to the limiting value $\log_2(6)$ for large interspecies repulsions.

Concerning the peaks, in agreement with what already discussed in section \ref{sec:CVP_Trimero}, one can observe that they broaden and move rightward upon an increase of ratio $T/(UN)$, a circumstance which mirrors the fact that the hopping amplitude $T$ has the twofold effect of smoothing the transitions and of delaying their occurrence. As regards the argument of the logarithm, \textit{six} is the number of classical configurations which minimize effective potential (\ref{eq:V_trimer}). Such configurations, depicted in the right panel of Fig. \ref{fig:Trimero_EE} and which are equal up to species labels exchange and up to cyclic permutations of the site indices, are quantum-mechanically reabsorbed in the formation of a unique ground state. The latter, in fact, exhibits a six-faced Schr\"{o}dinger-cat-like structure \cite{NoiSREP} whose explicit expression can be readily obtained properly generalizing equation (\ref{eq:Dimero_CAT}).

\subsection{Chaos and persistent demixing}
\label{sec:chaos}
So far, our discussion has been focused on the miscibility properties of the two species when the system is in its lowest energy state. A much broader panorama of physical phenomena emerges if one extends the analysis to excited states. In view of this complexity, which makes the scenario remarkably rich and branched, an exhaustive exploration of all possible dynamical regimes would go beyond the scope of this work. For this reason, we restrict our investigation to the dynamics of highly demixed configurations  \cite{Noi_NJP}. More specifically, we numerically simulate extended bundles of trajectories starting in the close vicinity of stationary states featuring a low degree of mixing. This is done in the semiclassical limit, where it is possible to replace second quantized operators $a_j$ and $b_j$ with local order parameters 
$$
    \sqrt{n_j}e^{i\phi_j} \approx a_j, \qquad \sqrt{m_j}e^{i\psi_j} \approx b_j,    
$$
complex numbers which, in turn, depend on local populations ($n_j$ and $m_j$) and local phases ($\phi_j$ and $\psi_j$).
After finding the semiclassical version of Hamiltonian (\ref{eq:Trimer_Hami}) and the motion equations that the latter induces, one can therefore indentify different classes of stationary configurations featuring  a trivial time evolution and a low value of the indicator 
\begin{equation}
    \label{eq:S_{mix}}
    S_{mix}=-\frac{1}{3}\sum_{j=1}^3\left[\frac{n_j}{n_j+m_j}\log\left(\frac{n_j}{n_j+m_j}\right) +  \frac{m_j}{n_j+m_j}\log\left(\frac{m_j}{n_j+m_j}\right) \right].
\end{equation}
The latter, which goes under the name of Entropy of Mixing, is commonly employed in Physical Chemistry to quantify the miscibility of chemical compounds \cite{Camesasca}, but its scope can be effectively extended to the realm of ultracold atoms, where it constitutes a reliable tool for monitoring the mixing of different condensed species. 

\begin{figure}
    \centering
    \includegraphics[width=1\columnwidth]{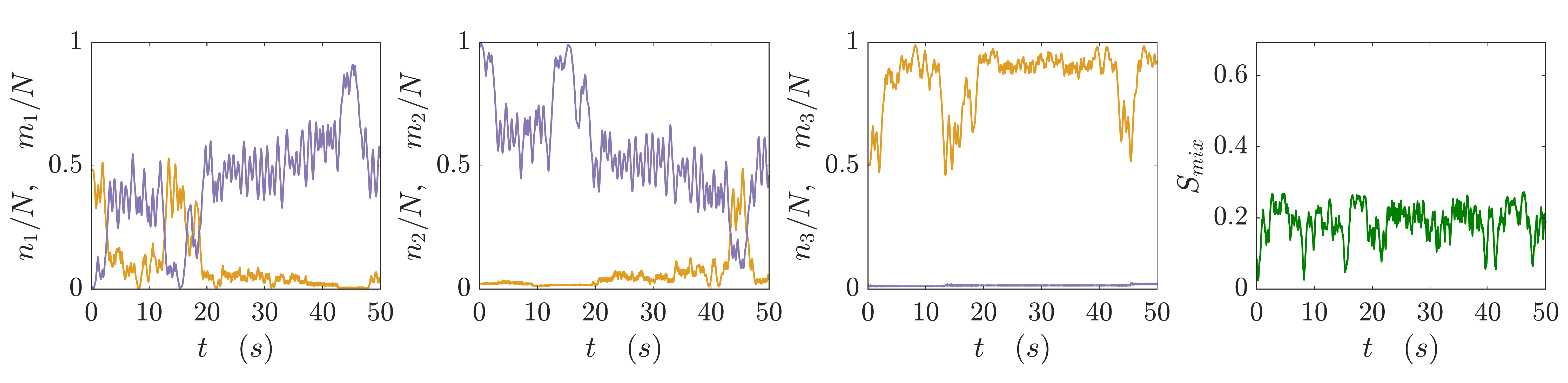}
    \caption{Time evolution of boson populations and of the relevant Entropy of mixing. The initial configuration is $\phi_1=0.1$, $\phi_2=-0.08$, $\phi_3=\pi$, $\psi_1=0$, $\psi_2=0.07$, $\psi_3=-0.05$, $n_1=24$, $n_2=0.6$, $n_3=25.4$, $m_1=0.58$, $m_2=49.04$, $m_3=0.38$. Despite the chaotic behaviour (the first Lyapunov exponent associated to this trajectory in phase space is $\lambda=2.12$), the Entropy of mixing remains always smaller than $0.28$ (note that the maximum possible value is $\log 2 \approx 0.69$). Model parameters $U=1$, $T=0.01$ $W=2.54$ and $N=M=50$ have been chosen.}
    \label{fig:Chaos}
\end{figure}

\begin{figure}
    \centering
    \includegraphics[width=0.7\columnwidth]{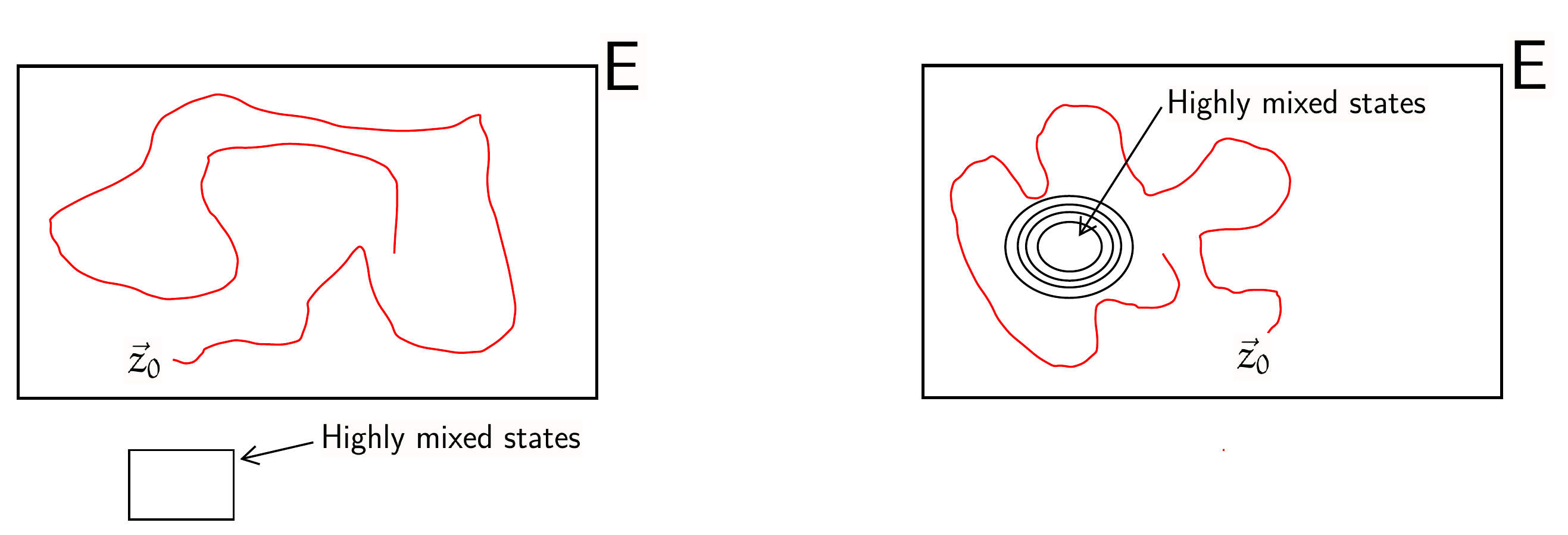}
    \caption{Pictorial representation of the two arguments behind the occurrence of persisting demixing in spite of chaos. $E$ represents the constant-energy manifold, $\Vec{z}_0$ a generic initial condition. \textbf{Left panel:} ``energetic argument" - highly mixed states lie outside the constant energy hypersurface and are therefore inaccessible, no matther the degree of chaoticity of the trajectory.  \textbf{Right panel:} ``dynamical argument" - highly mixed states, although being energetically accessible, are encapsulated within regular islands and cannot be visited by chaotic trajectories originating out of them.}
    \label{fig:Phase_Space}
\end{figure}

Interestingly, by exploiting this indicator, we have shown \cite{Noi_NJP} that many dynamical regimes feature persistent demixing in spite of a strong chaotic behaviour.  For example, Fig. \ref{fig:Chaos} clearly shows that the two species, despite their chaotic evolution, tend to exclude each other throughout the simulated dynamics. This kind of phenomenology, which has been evidenced by means of an explicit calculation of the first Lyapunov exponent and of indicator (\ref{eq:S_{mix}}) along extended sets of trajectories, can be interpreted either in terms of energy conservation or by means of dynamical considerations. Both these motivations, pictorially sketched in the two panels of Fig. \ref{fig:Phase_Space}, are based on the analysis of the constant energy hypersurface and of the orbits embedded therein.

Concerning the ``energetic argument", it is possible to show that the constant-energy manifold where the dynamics develops may not contain highly mixed states at all. In this circumstance, therefore, no matter how large the first Lyapunov exponent is, species mixing will never occur. As regards the ``dynamical argument", after recalling that the constant-energy hypersurface of low-dimensional systems features a coexistence of both regular and chaotic orbits \cite{Kandrup}, one can notice that, in the system under analysis, the former constitute regular islands where highly mixed states are encapsulated and where chaotic trajectories cannot enter. In this scenario, therefore, mixing is energetically possible, but actually forbidden by dynamical restrictions.      

\section{Conclusions and future perspectives}
In this work we have provided an overview on the rich phenomenology which distinguishes bosonic binary mixtures in low-dimensional systems. In section \ref{sec:Dimero}, we started our discussion with the two-well (dimer) system, the simplest possible geometry whose spatially-fragmented character has a remarkable impact on the miscibility properties of the species. By means of a semiclassical approach, we have evidenced the occurrence of a single mixing-demixing transition, from a fully mixed to a completely demixed phase, providing the analytic expression of boson populations for all the possible values of the control parameter $W/U$. To support the semiclassical analysis, we have also investigated genuinely quantum indicators, such as the energy spectrum and the entanglement entropy, highlighting their singular behaviour at the critical point. In section \ref{sec:trimero}, we turned to the trimer. The higher complexity of its geometry (given by the presence of an additional well) is mirrored by a more sophisticated phase-separation mechanism. In this context, in fact, an additional intermediate phase, neither fully mixed nor completely demixed, can be identified. The presence of two critical points is confirmed by the doubly-critical character of the energetic fingerprint and of the entanglement between the species. Interestingly, as in the dimer system, all the indicators that we have considered show that the hopping amplitude $T$ has the twofold effect of smoothing the transitions and of delaying their occurrence. In the last part of the manuscript, in order to go beyond the ground state's physics, we have extended our analysis to excited states, investigating the interplay between persistent demixing and chaotic behaviour. 

We hope to have made it clear that phase separation phenomena in low-dimensional systems constitute an extraordinary rich panorama and their analysis can touch on several different aspects, ranging from entanglement properties to quantum dynamics. The work that has been done so far has shined light just on those systems featuring the most elementary geometries (dimer and trimer). We therefore aim to extend it to more complex geometries where, presumably, multiple intermediate phases will come into play. Moreover, we would like to generalize our analytic and numerical computations to the case of non-symmetric species, i.e. $U_a\neq U_b$, $T_a\neq T_b$ and $N\neq M$, a situation where the presence of additional control parameters should offer an even more intriguing physics. Current work is focused on the study of \textit{attractive} interspecies interactions, which constitutes a realm on its own, as they lead to the emergence of phases featuring supermixed states rather than phase separation. We will report on this topic in a separate, dedicated work.

\vfill
\clearpage

\section*{References}
\providecommand{\newblock}{}


\end{document}